# Anomalous Weight Behavior in $YBa_2Cu_3O_7$ Compounds at Low Temperature

May 30, 1997

Frederic N. Rounds

## Abstract

$YBa_2Cu_3O_7$ high temperature superconductor samples were weighed on an electronic balance during a warming cycle beginning at $77^0K$. The experiment was configured so that the $YBa_2Cu_3O_7$ material was weighed along with a magnet, a target mass, and liquid Nitrogen coolant. The weights were captured during Nitrogen evaporation. Results indicated unexpected variations in the system weight that appear as a function of temperature and possibly other parameters.

## I. Introduction

Dr. Eugene Podkletnov reports[1,2] an experiment wherein a rotating, supercooled $YBa_2Cu_3O_{7-x}$ disk apparently acts as a shield between the Earth's gravitational field and a Silicon Dioxide sample suspended from the arm of an electronic balance. Sample weight changes are reported ranging from .05% to .3% depending on the velocity of rotation of the superconductor. During the period of time the weight effects were recorded the superconductor had a temperature range between 20K to 70K. The superconductor was suspended over a toroidal solenoid and was in a state of levitation due to the Meissner Effect. A varying magnetic field was placed around the superconductor and the frequency was varied between 50 HZ and $10^6$ HZ. The greatest weight change occurred at the highest frequency.  However, the weight reductions apparently occurred even when the rotating field was turned off and the superconductor stopped rotating.  Further anomalous weight behavior was reported[2] using Podkletnov's experimental design, but with significant simplifications.

The experimental designs reported to date have basically followed the Podkletnov approach.  That is  using a weight suspended over the top of the superconductor (SC), and the weight being connected to a balance.  Essentially we can view the weight and the SC  as two separate systems.  In such an experimental arrangement determining if "gravitational shielding" is at work as opposed to an action-reaction force, such as magnetic or electric fields, is very difficult.  Therefore, the experimental approach described herein creates a system in which the action-reaction components cancel each other.

The experimental question is whether the $YBa_2Cu_3O_7$ SC material acts as a gravitational shield in that mass situated above the SC loses weight while the SC is in the supercooled state.  It must be noted that very little theory exists[4,5,6,7] that predicts a gravitational interaction with SC's at the macroscopic level.  Most assuredly no engineering formulas exist that aid in system design.

## II.    Experimental  Design

The experimental approach reported herein has two phases:

Phase 1.   The determination of weight anomalies in a system consisting of magnet, coolant, SC, and target mass.

Phase 2.   The determination of whether variations in the target mass produces a proportionate variation in system weight.

Figure (1) shows the test model used in phase 1.  It consists of a 50 cc plastic prescription medicine bottle.  A rare earth Cobalt disk magnet (approximately .5 Tesla, Edmund Scientific part number C52, 867) is attached 1.9 cm below a 2.54 cm-by-.31 cm SC (Edmund Scientific part number C37, 446) disk. The magnet is held in place by sytrofoam and plastic resin is used as a platform for the SC.  A small plastic bead is used to provide air space between the SC and the resin surface.  The SC, bead, and resin platform are glued together.  A number 2 rubber stopper is glued to the bottle's cap to act as a target.  This target mass is approximately 10 grams.   Four .5 cm holes are drilled in the bottle just above the top surface of the SC.  These holes allow liquid Nitrogen(LN2) to enter and act as the coolant during the weighing.  The bottle plus magnet, SC, target, and LN2 will be heretofore designated as the *system*.  Not having any engineering guidelines for the construction of this test model, the criticality of the dimensions provided are uncertain.

The experimental methodology is merely to cool the SC to $77^0K$ using an LN2 bath.  Some LN2 enters the bottle through the holes.  The system is then placed on a sensitive digital electronic balance. In the case of this experiment a Sartorius Model 1207-MP is used. It has a readability of .1 mg and it updates its reading every second. The balance has glass sliding doors at the top and both sides, so the model can be completely enclosed for more accurate weighing. A six-ounce styrofoam cup separates the balance's pan from the system. This distance seems sufficient to prevent the magnet from interacting with the balance's electro-mechanics. However, balances are subject to magnetic influences and this is an area of noise that could not be completely eliminated with certainty.

As the system sits on the balance the LN2 is evaporating and, therefore, mass is leaving the system and the balance reads a continual weight decrease over time. The balance's digital readout was videotaped during the full span of time that LN2 is evaporating and beyond to assure that weights are gathered after the SC warms passed its critical temperature, $T_c$, of approximately $90^0 K$. The video camera uses 8mm tape and contains a time and date stamp so the time down to 1-second intervals can be recorded along with the weight readings. $T_c$ can vary by several degrees from one piece of $YBa_2Cu_2O_7$ to another. The exact value of $T_c$ was not determined during the course of these experiments. The LN2 in the system appeared completely evaporated after about 2 minutes.

After each weighing cycle the videotaped readings are transcribed to an Microsoft Excel Version 5 spreadsheet. Both the time and the corresponding weight are placed on the spreadsheet. The graphs contained herein were generated using Excel. The raw data plots are presented along with a 4-point average plot. The 4-point average serves to smoothe out the low level noise, but does not disguise the anomalous weight behavior.

The test model in Figure (2) was used for Phase 2 testing. The magnet and SC are the same as in Phase 1. The magnet, however, is glued directly to the SC, and then this complex is glued to the bottom of the bottle. A nylon bolt is attached to the bottle's cap. 2.38 cm X .31 cm rubber washers are then bolted in as required to vary the weight. To add more weight two brass washers are used in place of the washers. This arrangement creates a rigid, but modifiable target.

## III. Predicted Results

In this experimental design, if the SC acts as a gravitational shield, then the system should display a gain in weight as it warms passed $T_c$. This gain in weight is expected because when the test model is placed on the balance, it is already in the supercooled state and, therefore, shielding should be occurring. In essence the system is lighter when it is placed on the balanced and in theory we would assume it would gain weight as $T_c$ is passed. The weight vs. time curve should show an abrupt slope change at the point of transition and then a continuation with the previous slope characterized by evaporating LN2.

**IV.  Phase 1 Results**

Four(4) trials were performed during phase 1.  Two trials were with the SC.  The third and fourth trials used rubber and brass controls respectively instead of a SC; yet, no other features were changed in the system.

Trial 1 results are shown in Figure (3).  Using the gravitational mass scale as shown we can see a high degree of linearity in the evaporation curve until a sharp slope change occurs at around 46 seconds.  This linearity is apparent down to the centigram level.  The sharp slope change is unusual, but can be easily explained as being the point in which the LN2 has completely evaporated and the balance's sampling rate is not fine enough to catch transitional points.

Nothing particularly unexpected appears in the weight vs. time curve. But, the linearity allows us to perform some statistical analysis that could reveal effects to a much lower level.  First, we make the assumption that $T_c$ occurs somewhere between 0 and 45 seconds.   Second, we estimate the actual evaporation curve as given in Figure (3), by fitting a line to the actual data. This is done simply by using the slope intercept formula with the first weight at time 0 and the second weight at time 45. The intercept is the point with the weight corresponding to time zero(0).

We know that a curve in two dimensions can be represented by the following formula:

$w(t) = l(t) + n(t)$, where w is weight, l is the linear part of the curve, n is the nonlinear, and t is time.

Therefore,

$n(t) = w(t) - l(t)$.

Letting $l(t)$ be the linear estimate, $n(t)$ represents the nonlinear behavior of $w(t)$ with respect to this line of estimation. $n(t)$ measures the deviation of weight from the linear. If $l(t)$ is very close to $w(t)$, then $n(t)$ will display very minute variations in the weight.

$n(t)$ is plotted in Figure (4). Note the linearity is so tight between $l(t)$ and $w(t)$ that $n(t)$'s scale ranges only 20 mg. We can easily see effects at the mg level. We can see the obvious 5 mg increase in weight beginning at about 16 seconds. The target mass weighed about 10 grams. Hence, we observe a .05% weight increase. This percentage is what Podkletnov noted as the lower level of observations he had made in his initial work.[1] This increase seems to evolve over a period of about 8 seconds. Then a sharp decent occurs to the original slope of the "pre-event" curve.

The same experiment was repeated in trial 2 to see if the results could be replicated. Figure (5) shows again the high linearity during the first 45 seconds. Figure (6) uses the linear estimation technique as in trial 1. Note the appearance of the weight increase beginning at about 25 seconds and continuing to about 33 seconds, an 8-second cycle. Also, this weight increase is again at about the .05% level. The fact that it occurs later in the warming period is an interesting variation from trial 1. Also, the concavity of the chart is concave down, rather than up as in trial 1.

One of the significant challenges in this experimental arena is finding the right combination of controls to narrow the possible explanations for the results seen in trials 1 and 2. Possibly the curves merely display the normal evaporation behavior of LN2. Or, possibly the magnet is interacting with the balance. Some insight can be gained by using a control sample material instead of the SC. Hence, trials 3 and 4 used rubber and brass respectively.

Figure (7) shows the evaporation curve for rubber. Note the linearity again, so we were immediately able to apply the linear estimate technique and plot the results shown in Figure 8. At the same mg scale as in previous trials no obvious and anomalous weight effects are observed.

The experiment was repeated using a brass control sample in trial 4. The weight vs. time curve is shown in Figure (9). The results here are a bit more nonlinear. However, we still applied the same linear estimation technique where the results are displayed in Figure (10). Again, we see no obvious weight anomalies.

**V.  Phase 2 Results**

Recall that in Phase 2 the test model was redesigned. The major change was actually putting the magnet much closer to the SC. This had the effect of increasing the local magnetic field entering the SC by a factor of over 1000. The Sample SC has a critical magnetic field much lower than .5 tesla. Therefore, the SC was being over-saturated by the magnetic field. The SC, therefore, could not achieve superconductivity. Over-saturating the SC was not intentional and it was discovered during the Phase 2 experimental calibrations and in conversations with the SC's manufacturer. The calibration performed was merely checking the presence of the Meissner Effect at LN2 temperature, $77^0K$. The Meissner Effect was never achieved with this configuration.

However, the trials performed in Phase 2 offered some interesting and surprising effects. In trial 1 of Phase 2 we use a 13 gram target weight. Note the subtle slope aberration at around 17 seconds in the weight vs. time curve, Figure (11). The slope decreases for a short burst then continues on with its previous slope. Using the linear estimation, Figure (12), we see that at the 17-second point what might be interpreted as a very subtle increase in weight, but this could be merely judged as noise. Figure (11) shows an obvious decrease in slope, which would indicate a small increase in system weight. Of course, these anomalies could easily be relegated to noise effects. Figure (13) displays a bar chart of the slopes from 7 seconds to 27 seconds of the trial run. The slope is simply $Weight_i - Weight_{i+1}$. Note that at 17 seconds the slope is markedly less than all others in the group. This decrease in slope can be explained by a system weight increase of .05% to approximately .1%.

Trial 2 increases the target mass to 27 grams using two brass washers. Figure (14) shows the weight vs. time curve and again we see the slope anomaly appear at 16 seconds. The linear estimation curve is shown in Figure (15) and the increase is clearly seen at the 16-second point, representing a weight increase of approximately .05% to .1% of the target weight. Doubling the target weight seems to produce a proportionate increase in system weight during warming.

**VI. Analysis of the Experiment**

The experimental data to date is very difficult to interpret. To assume that gravitational effects are at work is premature. Other possibilities exist and will be discussed below.

1. **Thermal conductivity changes in the superconductor**. If the thermal conductivity of the superconductor changes around the critical temperature, then we would see changes in the LN2 evaporation rate, which could account for some of the observed results. However, these effects could be ruled out because the thermal conductivity of this particular type of SC actually begins to increase at $T_c$, and peaks at about $50^0$ K.[5] The increase is small from approximately 3.5 to 4.0 watts/meter/ K degree. The conductivity is very flat up to around $150^0$K. So, the smallness of the change in thermal conductivity is not expected to produce the observed results. Yet, even more is the fact that the thermal conductivity increases. This would cause the evaporation rate to increase, not decrease. Therefore, such a change in thermal conductivity would be observed as a decrease in system weight, not an increase.

2. **External magnetic fields**. Such fields are the most viable candidates for the observed results and the most difficult to rule out. The experimental surroundings were not shielded against electromagnetic fields. The fact that the rubber and brass trials showed none of the effects observed with the SC trials, indicates that the magnet attached to the system did not contribute by itself to the anomalous effects. However, the controls do not rule out magnetic effects brought on by the superconducting material, even though the SC does not achieve superconductivity in Phase 2. The observation that the proportion of weight increase remains constant even as the target weight is doubled also indirectly diminishes the external magnetic field possibility, and contributes some validity to a gravitational shielding explanation. However, further testing is necessary to rule out external field influences.

3. **Effects within the Electronic Balance**. These effects are good candidates and must be ruled out by repeating the experiment using other equipment. The assumption in these experiments is that effects within the balance would occur at random points and would not occur in the same places or in the same way during successive trials. Therefore, balance effects would not explain the repeatable results obtained in these experiments.

4. **Random Noise**. We can quickly rule out random effects because of the fact that the experimental behavior showed consistent and similar results during each trial execution.

5. **Atmospheric effects**. Because LN2 is evaporating inside the closed balance chamber, significant thermal effects are occurring. Convection currents would be prevalent and the balance may be effected by the temperature changes. The assumption in these experiments is that atmospheric effects would behave in a random manner and would show up differently during the experimental replications. However, ideally this experiment could be greatly improved by performing it in a vacuum.

6. **Rapid condensation of water vapor**. These effects are assumed to behave randomly.

## VI. Hypothesis

To provide a hypothesis as to the cause of these observed phenomena is tempting. The effects seen by varying the target mass appears to point to a gravitational connection. However, the fact that in the experiment superconduction was prevented by an over-saturating magnetic field would lead one to guess that the material displays some temperature dependent critical points in its own gravitational interaction, which are possibly independent of the electronic effects causing superconduction. The critical temperature for the gravitational interaction may be somewhat different than that of superconduction. No viable theory at presents explains or predicts the phenomena observed in these experiments. Obviously a great deal more work is required with rigorous examination and replication within the scientific community.

Since these effects were observed at the macroscopic level, a useful endeavor might be to describe the phenomena using a Newtonian perspective.

Minimally we observed in these experiments a weight anomaly at the .05% level. Let

$$\text{Force} = GM_1 M_{target}/R^2$$

where $G = 6.67 \times 10^{-11}$ nt-m$^2$/kg$^2$,

$M_1$ = mass of the Earth at $5.98 \times 10^{24}$ kg,

$R$ = the radius of the Earth at $6.37 \times 10^6$ meters,

$M_{target}$ = the mass of the experimental target, .01 kg for the first set of trials.

$GM_1/R^2 = 9.86$ using the values provided.

Hence, the force between the target mass and the earth is

$$9.86 \times .01 = .0986 \text{ nts.}$$

The anomaly appeared as a decrease in weight of .05%, or $4.93 \times 10^{-5}$ nts.[9] The question now arises regarding what is really changing to produce the weight anomaly: target mass, Earth mass, the gravitational field, or some other parameter(s). To assume that the masses have changed would immediately open additional questions regarding whether inertial mass changed equally. Is the Principle of Equivalence violated? Another perspective would be to assume that the gravitational field is altered locally above the SC and the masses are left unchanged. This would imply the gravitational constant G changes at some point around the SC's critical temperature.

G has dimension Length$(L)^3$/(Mass(M) x Time$(T)^2$). If we assume that mass does not change, then we are left with length and/or time. For simplicity sake, assume that time does not change. Therefore, we can argue that if the field constant G changes it is a result of a change in length measure. Or, we can establish a very restricted form of relativity within the domain of superconductivity.

Since all variables. except length, are considered invariant under a transformation that occurs at the SC's critical temperature($T_c$), we can immediately write a relation between reference frames above and below $T_c$.

$$L'/L = (F'/F)^{1/3},$$ where L' and F' are the length

and force measures as observed below $T_c$.

In the case of trial 1 with a 10 gram target mass, F'/F = .999995. Therefore, L'/L = .999998. Or, L' = .999998L.

We can use the information determined above as a form of boundary condition in determining a metric transformation equation. The behavior of the weight in these experiments can be described as shown in Figure (16).

Note the step near the critical temperature of approximately $90^0$K. We know that superconducting materials behave electrically as shown in the figure in that electrical resistance also drops to zero as a step. The following equation is a good representation of the experimental behavior:

$$\text{(1)} \qquad L' = L\,(1 + \beta e^{(1 - T^n/T_c^n)})^{-1}$$

where T is the temperature in degrees Kelvin. $T_c$ is the critical temperature. ß is the appropriate scale factor determined by boundary conditions. n is a real number that is determined by the bandwidth around $T_c$. n and ß need to be determined experimentally because they seem to be material dependent. In Figure (16) for this discussion we let ß = .999998, $T_c$ = 90K, and n = 10. n does not need to be a whole number.

    Using Equation (1) we can begin asking questions similar to those related to Special Relativity. What do observers in one reference frame see happening to events occurring in another reference frame? In the case of these experiments we are speculating that the reference frames are related as a function of temperature, such as Equation (1). The laws of physics transform invariantly because Equation (1) merely behaves as a constant in derivatives of L and L' with respect to time. The distance metric also preserves its form, so Maxwell's Equations and General Relativity transform invariantly. The conclusion of this speculation is the observers in the above-$T_c$ frame will see measured differences in physical phenomena, such as weight, occurring in the below-$T_c$ frame.

## VII. Conclusions

A great deal more work is required to determine the validity of this anomalous weight behavior. The observed effects in this report and in others have been very small, but definitely on the macroscopic level and they are detectable with simple apparatus. Most high school laboratories could repeat the experiments described in this report. Replication of the experiment described in these pages is encouraged, but the following avenues of research also need further investigation:

- Test the effect of cascading magnets and SC's above one another. Would this increase the anomalous effects?
- Different configurations of magnetic fields, both static and varying need to be examined.
- Different types of SC materials need to be tested, such as Bismuth compounds and other varieties of high temperature SC's. Not all SC materials may behave as the one in these experiments.
- Further tests need to be made using increasing target mass.

# VIII. Acknowledgments

In doing this series of experiments I received great encouragement from four people from around the world. Mr. John Schnurer provided many insights into the behavior of the materials and has helped analyze the results to date. He also has observed this puzzling behavior through his own laboratory efforts. Dr. Giovanni Modanese has not only been encouraging and has helped me calibrate what possible theoretical basis exists for the observed behavior. I appreciate the help that Dr. David Noever has provided me during these difficult experimental trials. He has been willing and interested in looking at the results and helping with trying to narrow the various reasons for the anomalies. I am extremely grateful to Michael Adamson from NASA Ames Research Center for helping me with the laboratory equipment. He was very cooperative during the strange hours of these experiments. Lastly, I appreciate the comments received from Phillip Carpenter of the Department of Energy, Oak Ridge Operations relative to the experiment and this report.

# Figure 1.
# Test Object or Gravity Shielding Experiment
# Phase 1 Model

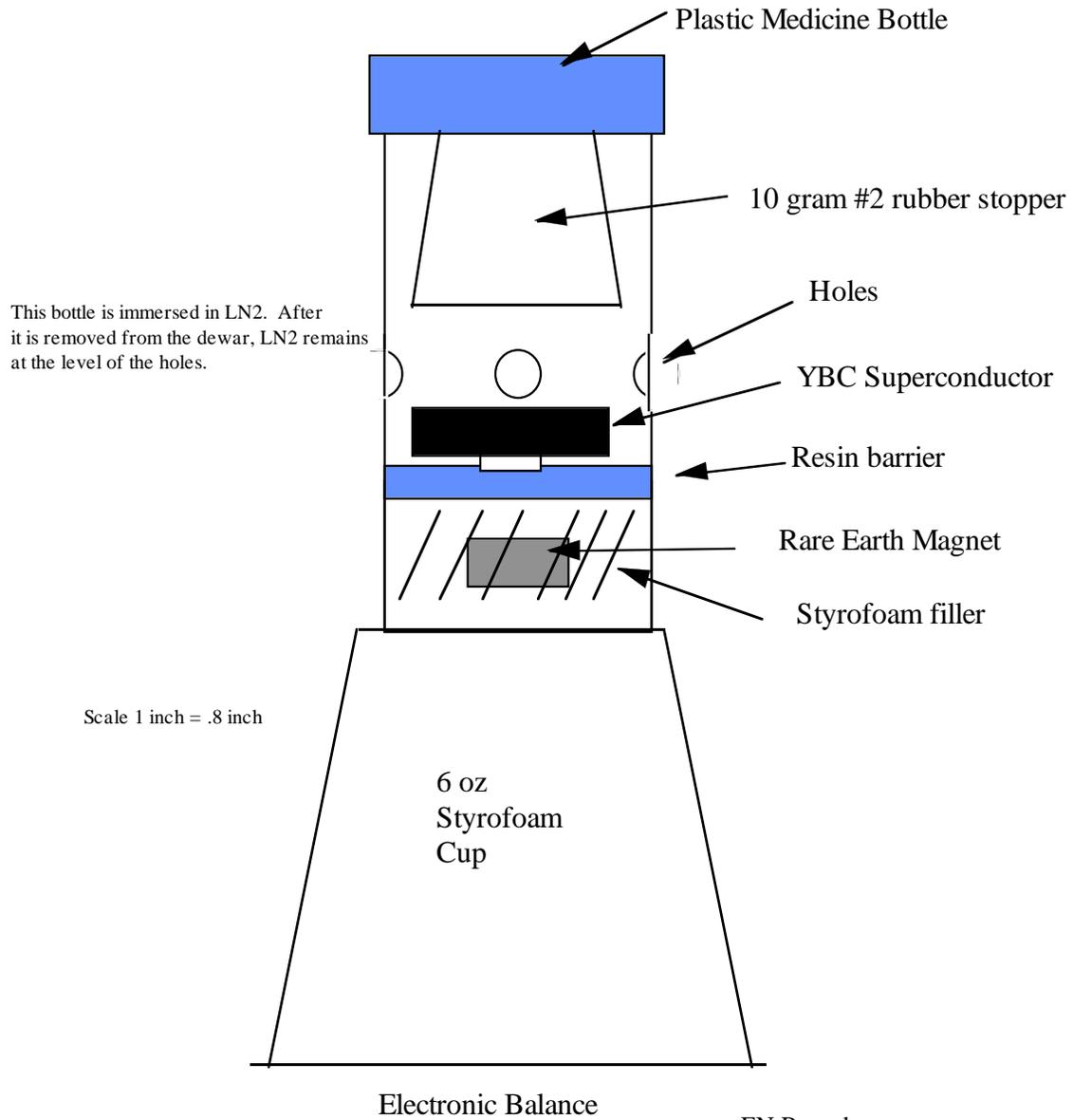

Scale 1 inch = .8 inch

FN Rounds
4/2/97

# Figure 2.
# Test Object or Gravity Shielding Experiment
# Phase 2 Model

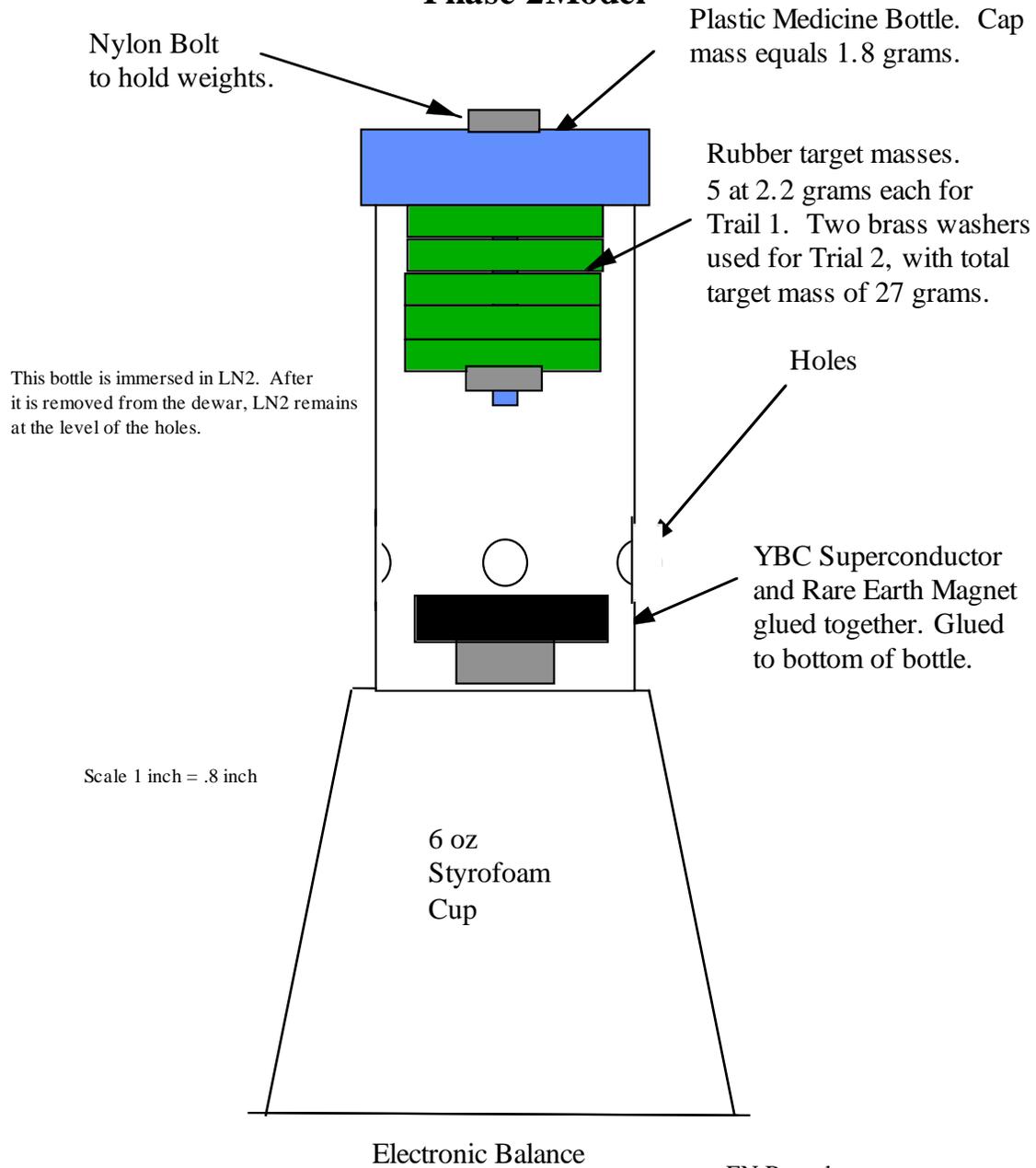

Nylon Bolt to hold weights.

Plastic Medicine Bottle. Cap mass equals 1.8 grams.

Rubber target masses. 5 at 2.2 grams each for Trail 1. Two brass washers used for Trial 2, with total target mass of 27 grams.

Holes

This bottle is immersed in LN2. After it is removed from the dewar, LN2 remains at the level of the holes.

YBC Superconductor and Rare Earth Magnet glued together. Glued to bottom of bottle.

Scale 1 inch = .8 inch

6 oz Styrofoam Cup

Electronic Balance

FN Rounds
4/2/97

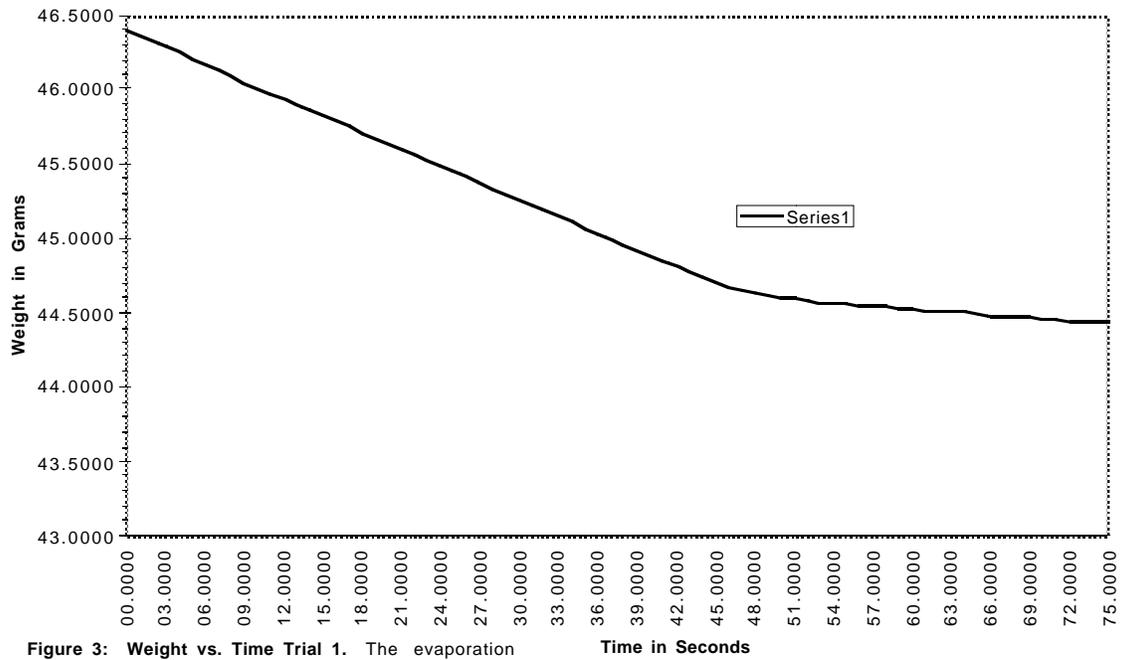

**Figure 3: Weight vs. Time Trial 1.** The evaporation curve is highly linear up to about 46 seconds. The slope breaks sharply then becomes linear again when the LN2 is completely evaporated.

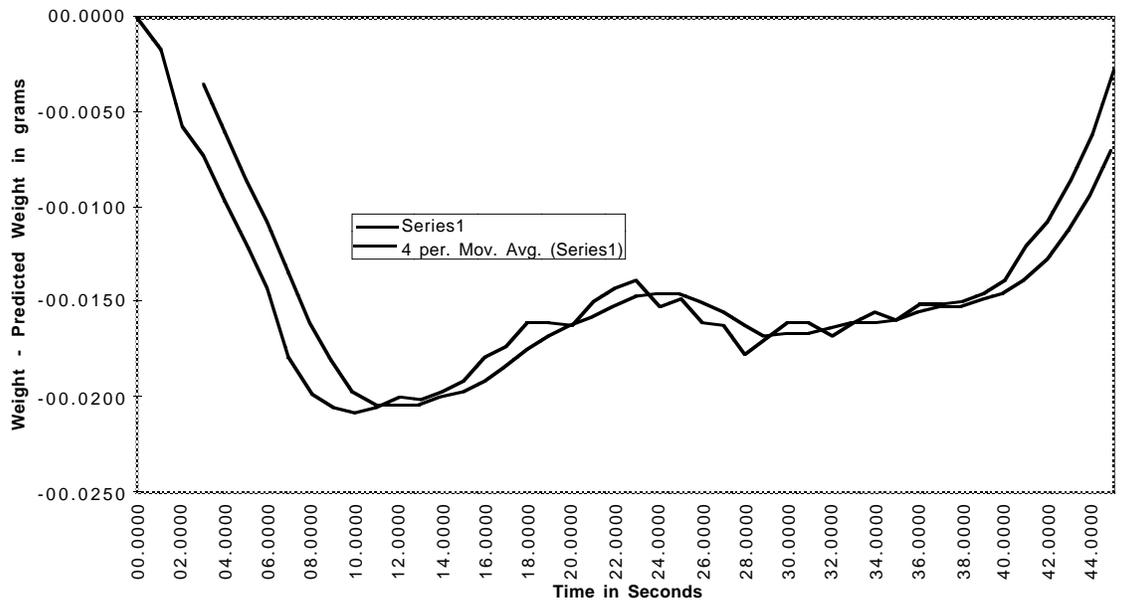

**Figure 4. Weight - Predicted Weight for Trial 1, 10-gram target.** Note the prominent weight increase beginning at about 16 seconds. Approximate system weight increase equals .05% of target weight.

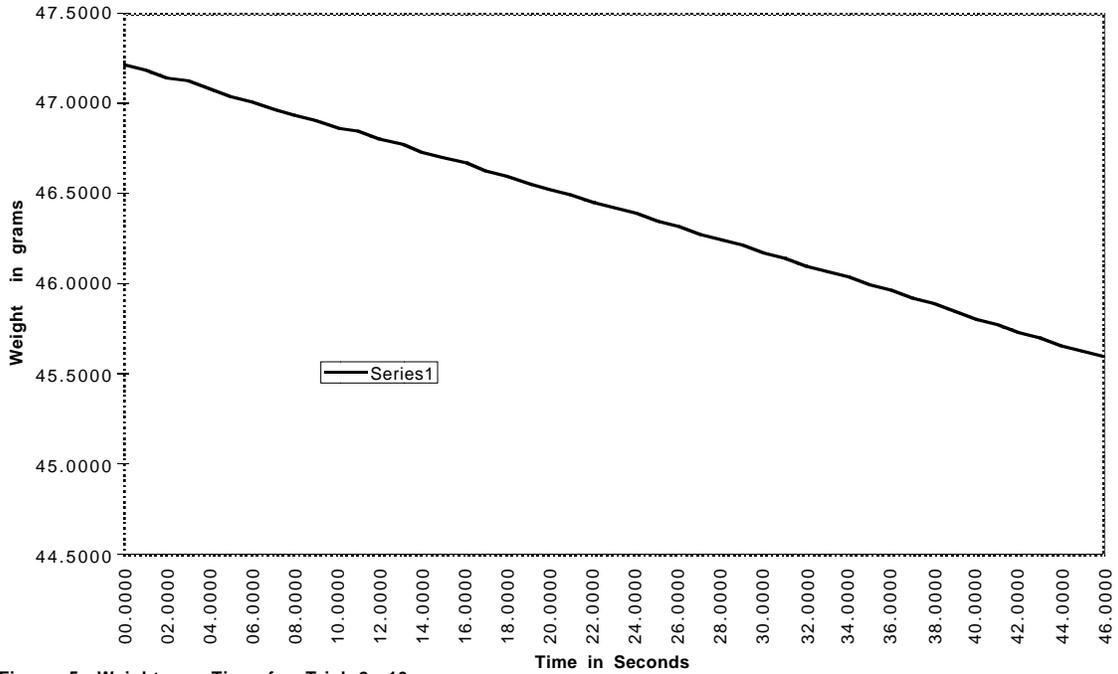

**Figure 5. Weight vs. Time for Trial 2, 10-gram Target.** Note again the high degree of linearity. This was the second trial using the Phase 1 test model.

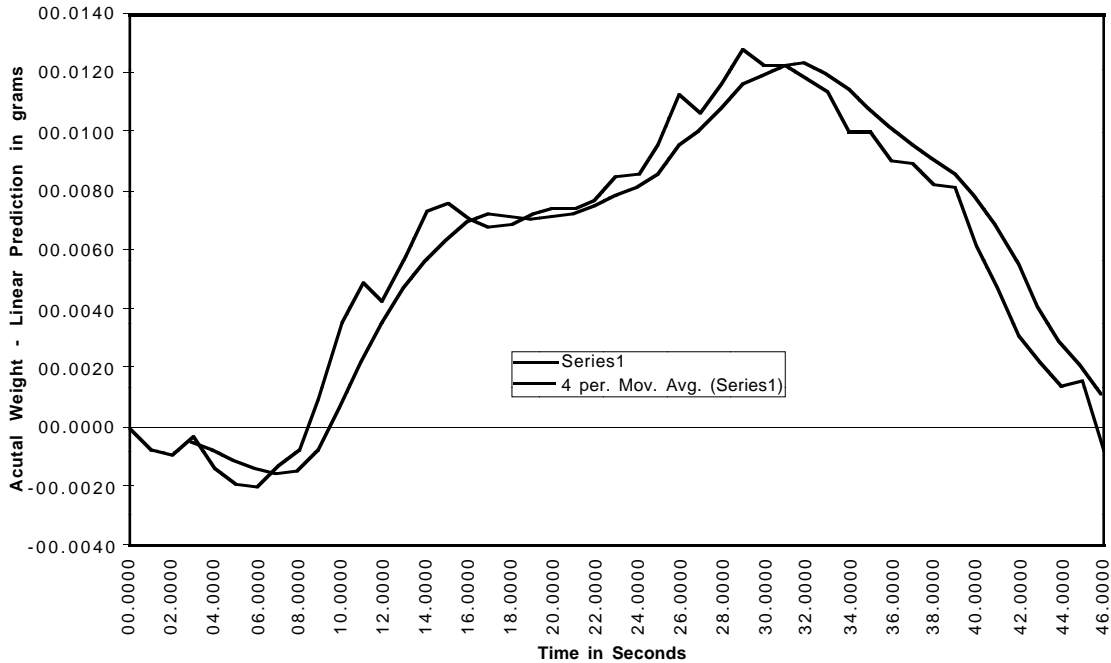

**Figure 6. Actual Weight - Linear Prediction Trial 2, 10-gram target weight.** The system weight anomalie appears at about 23 seconds which is later than the previous trial. The increase is again about .05% of the target weight.

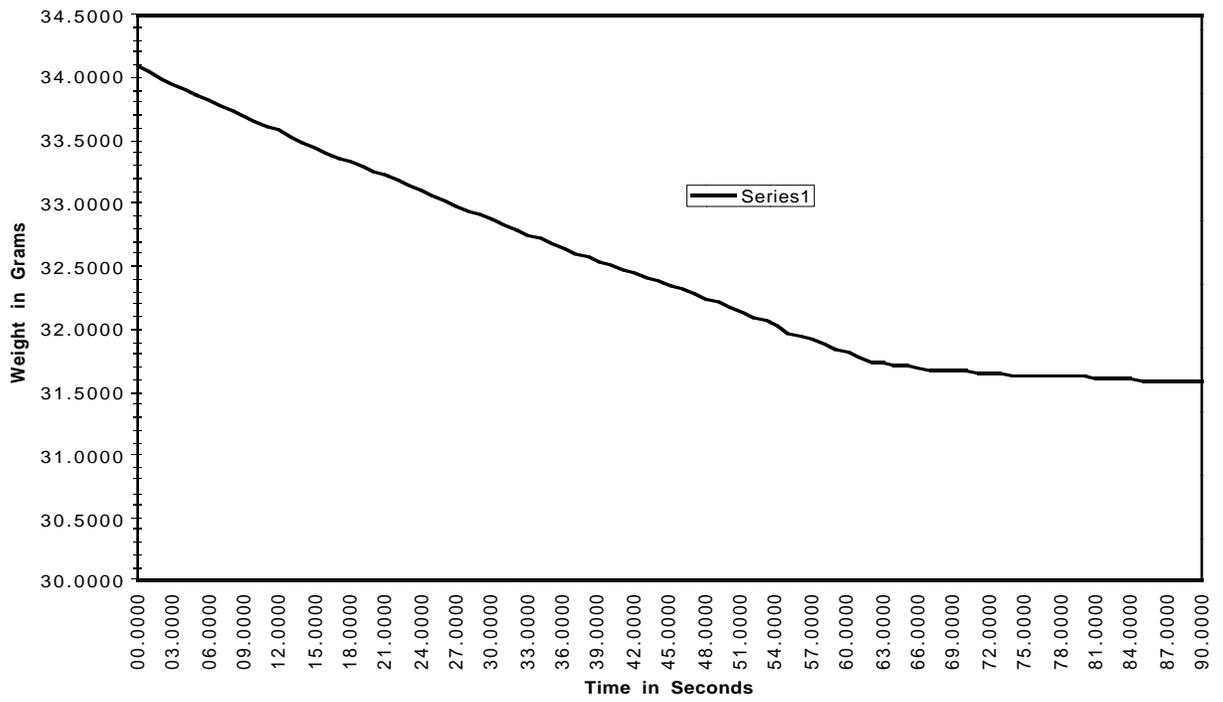

**Figure 7. Weight vs. Time for Rubber Control Sample, 10-gram target mass.** -
The evaporation is highly linear until LN2 is completely evaporated.

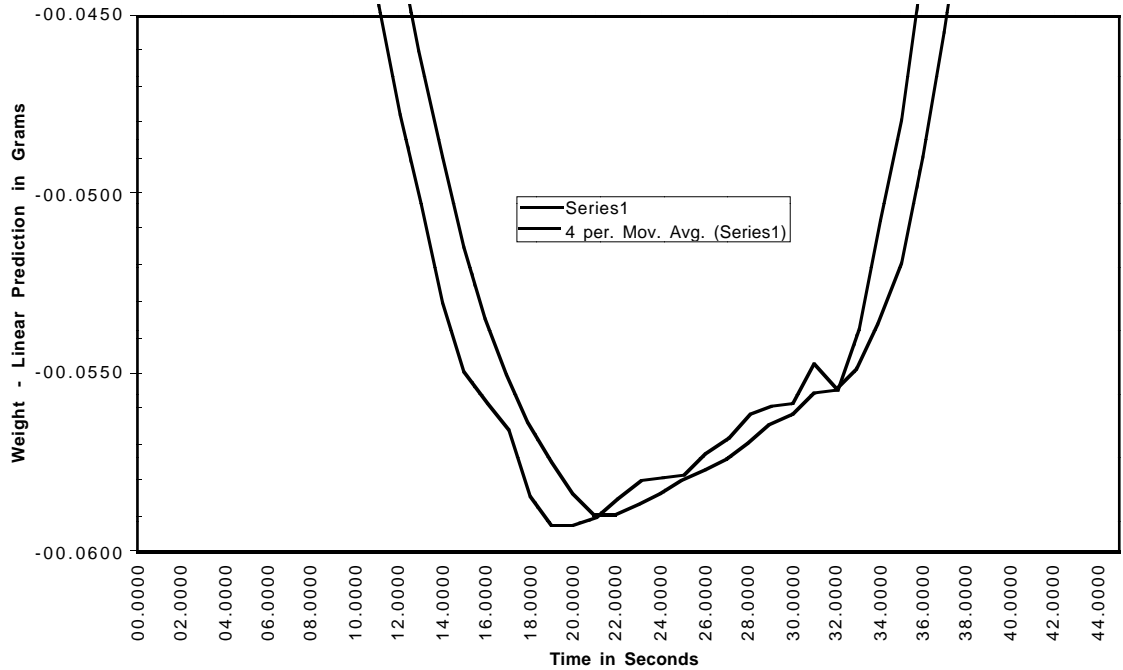

**Figure 8. Weight - Linear Prediction for Rubber Control Sample, 10-gram target mass.** No significant weight anomalies occur that can be distinguished from noise.

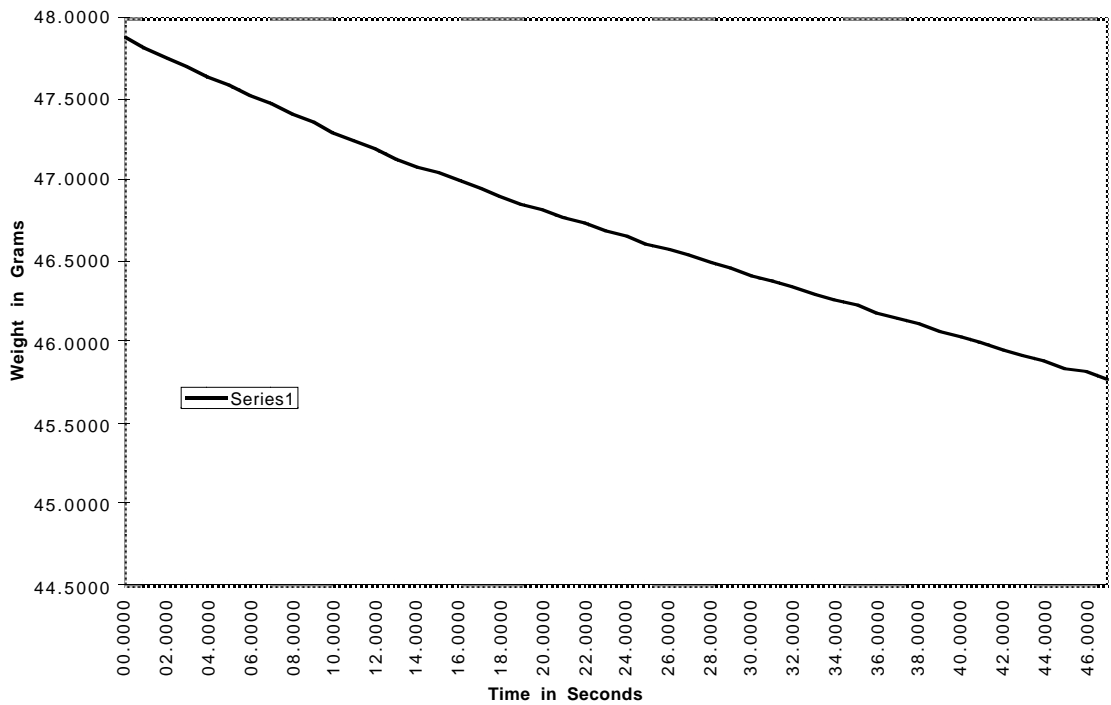

**Figure 9. Weight vs. Time for Brass Control Sample, 10-gram target.** The evaporation curve is more nonlinear.

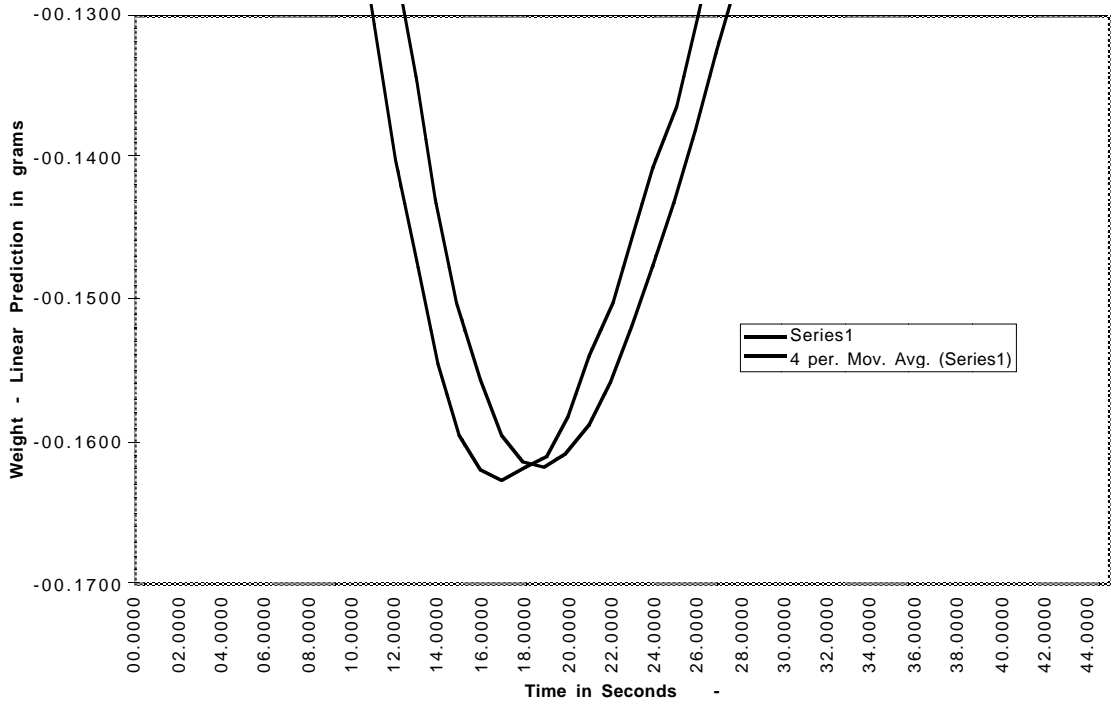

**Figure 10. Weight - Linear Predicted Weight for Brass Control Sample, 10-gram target mass.** The evaporation curves are more nonlinear, but weight anomalies can not be detected beyond the noise.

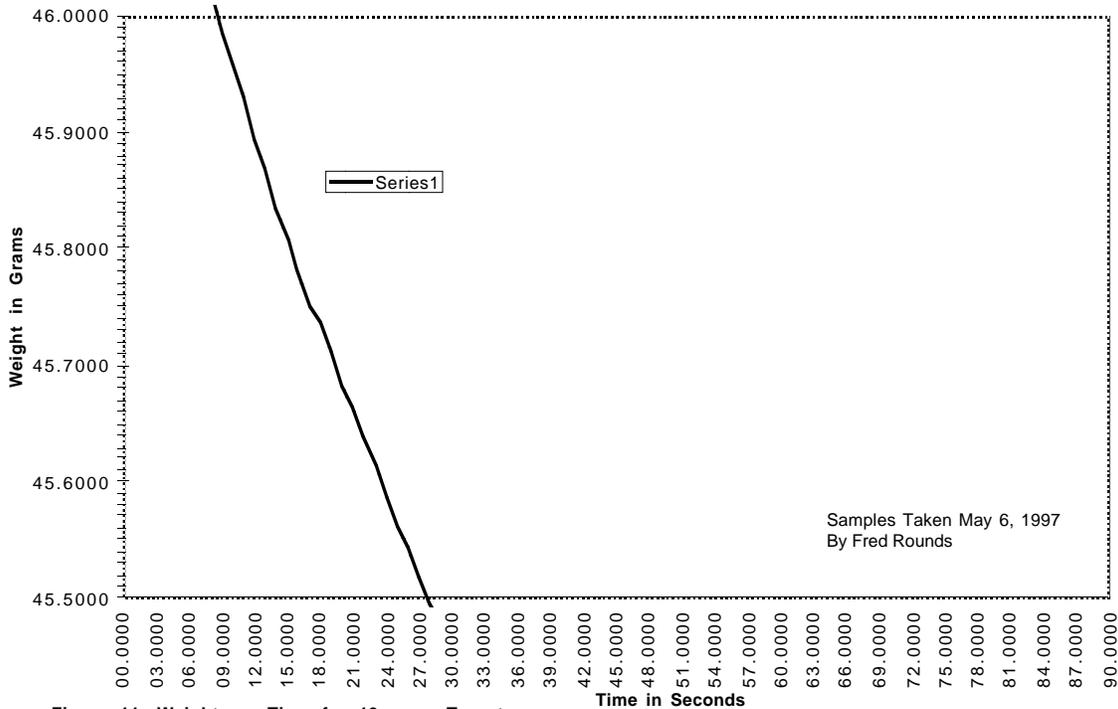

Samples Taken May 6, 1997
By Fred Rounds

**Figure 11. Weight vs. Time for 13 gram Target.** Note the anomaly at 17 seconds. This displays a instantaneous decrease in slope which points to an increased system weight.

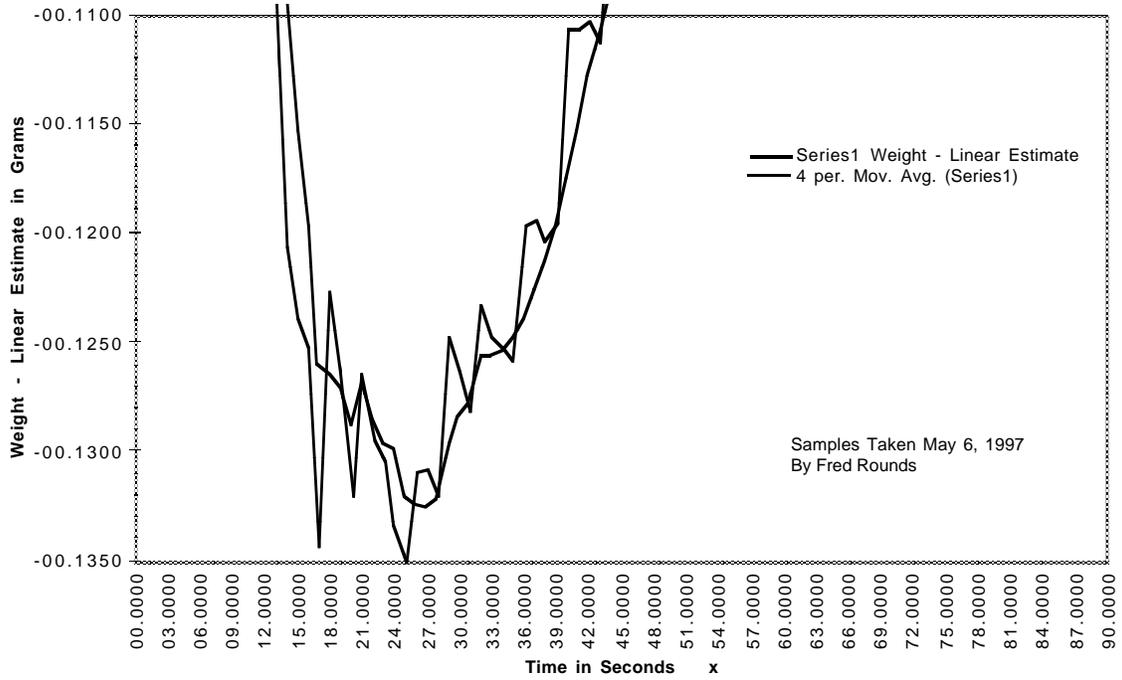

**Figure 12. Weight - Linear Estimate for 13 gram Target.** -
The anomaly is difficult to see but it occurs at 17 seconds.
See the bar chart for a clearer view.

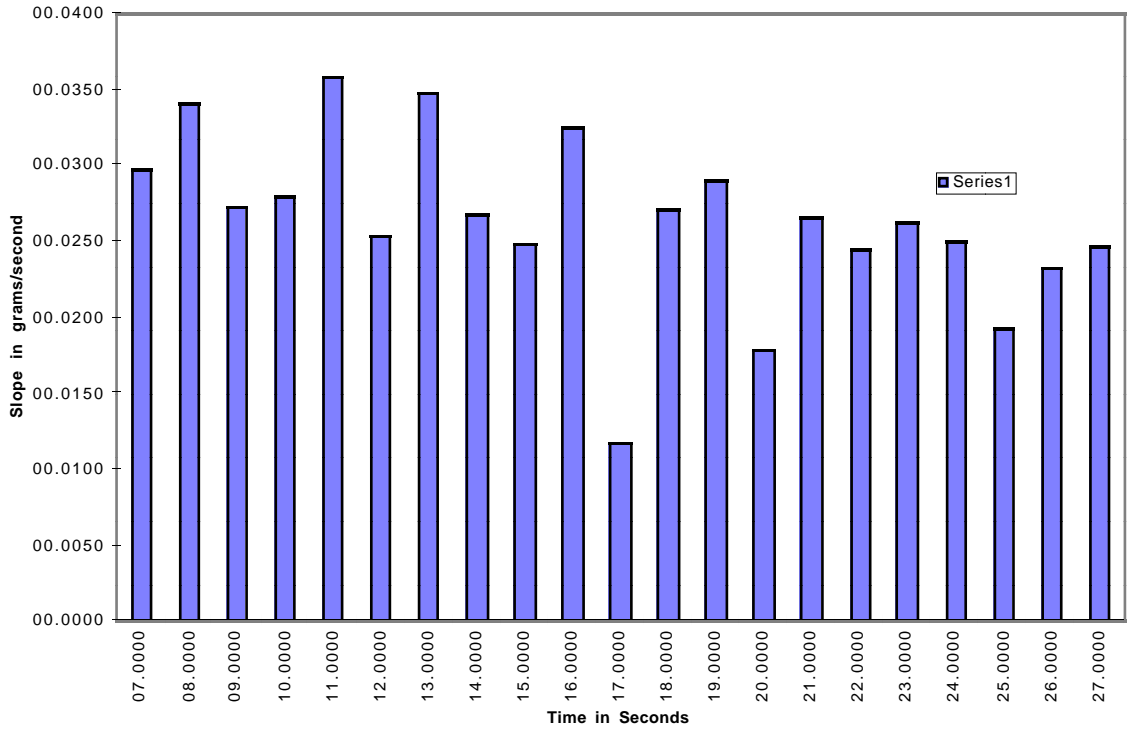

**Figure 13. Slope behavior of Trial with 13 Gram Target.** -
Note the significant drop in slop at 17 seconds. This
could represent a .05% to .1% increase in system weight.

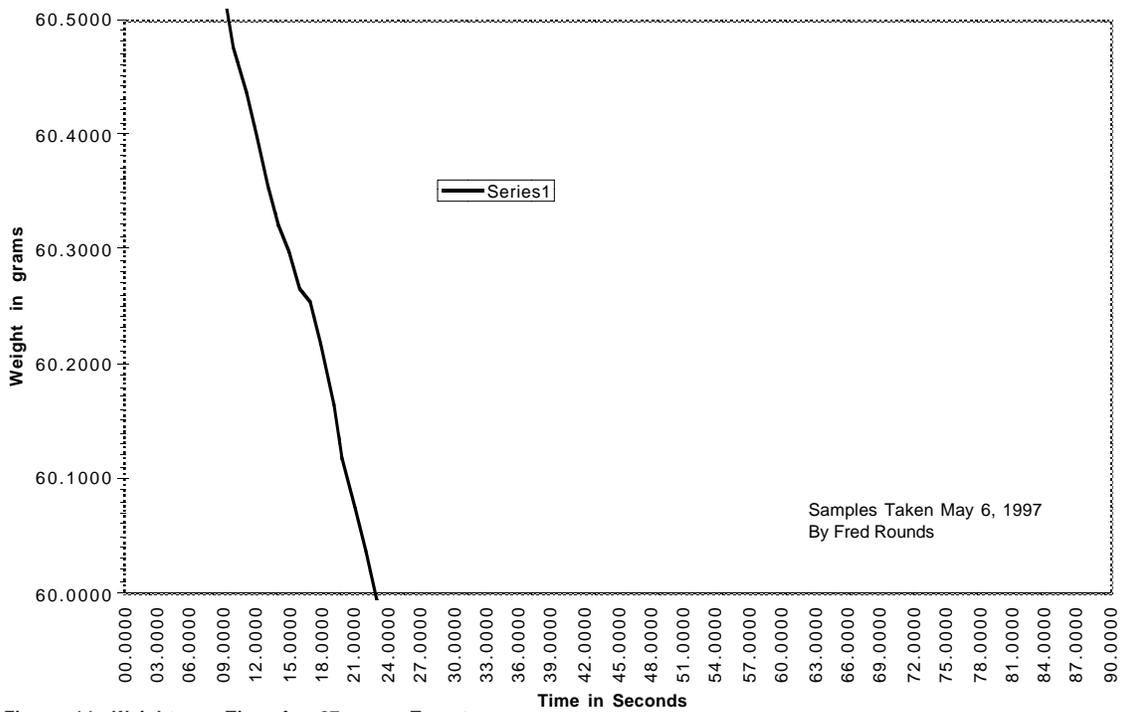

Samples Taken May 6, 1997
By Fred Rounds

**Figure 14. Weight vs. Time for 27 gram Target.** -
Note the anomaly at 16 seconds. It occurs at approximately
at the same time sequence as the 13-gram trial.

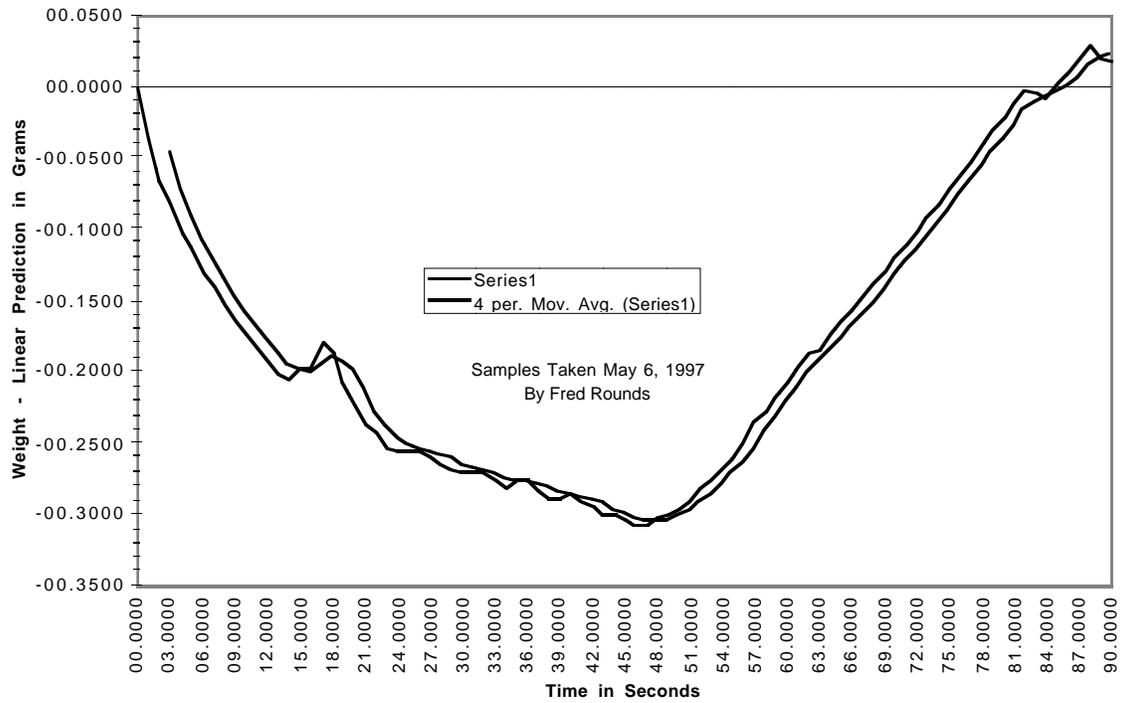

**Figure 15. Weight - Linear Estimate for 27 gram Target.** -
Note the obvious anomaly at 16 seconds. This represents
again system weight increase of from .05% to .1%.

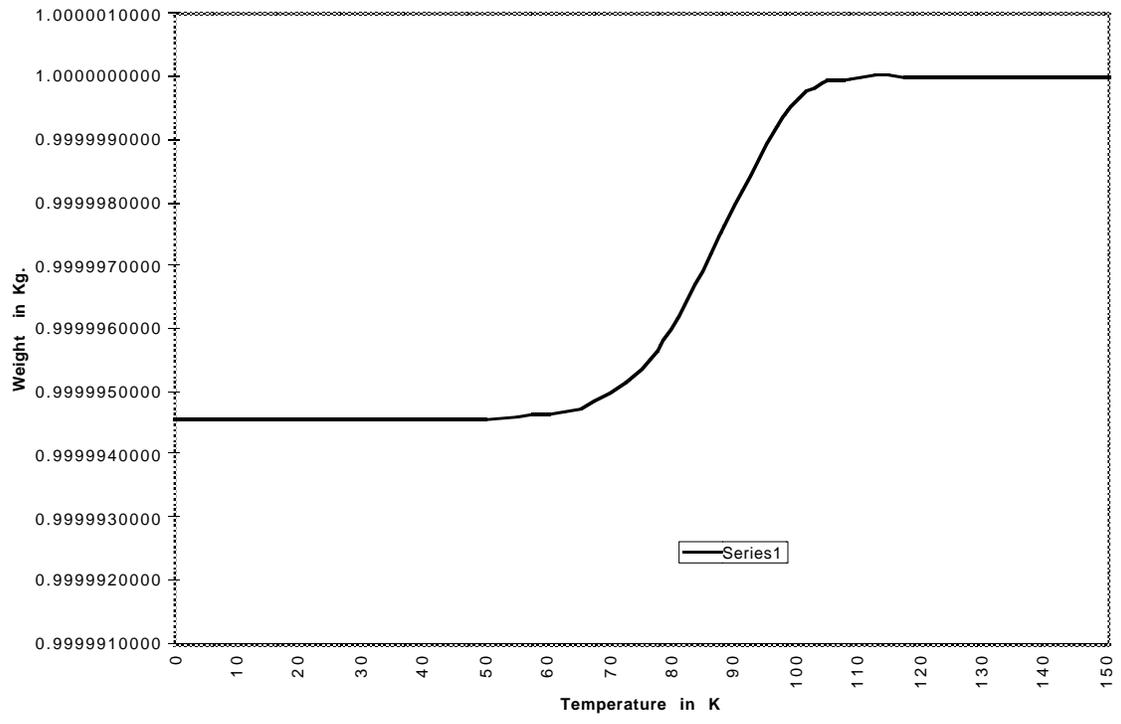

**Figure 16. Weight vs. Temperature. -**
The weight behaves as a step function
near the critical temperature.